\documentclass[aps,floatfix,showpacs,superscriptaddress,preprint]{revtex4}
\usepackage{graphicx,epsfig,epstopdf}
\usepackage{amssymb,amsmath,amsxtra,amsfonts,latexsym}
\usepackage{bm}
\usepackage{float}
\usepackage{longtable}
\usepackage{multirow}
\usepackage{booktabs}
\usepackage{array}
\usepackage{wrapfig}
\usepackage{color}
\usepackage{slashed}
\usepackage{bbm}
\usepackage{titlesec}
\usepackage [latin1]{inputenc}
\titleformat{\section}{\large\bfseries}{\thesection}{1em}{}

\newcommand{\bea}{\begin{eqnarray}}
\newcommand{\ena}{\end{eqnarray}}
\newcommand{\be}{\begin{equation}}
\newcommand{\en}{\end{equation}}
\newcommand{\nn}{\nonumber\\}
\newcommand{\ed}{\end{document}}
\newcommand{\Tr}{\mbox{\rm{tr}}}

\newcommand{\Jpsi}{\ensuremath{J\!/\!\psi}}

\begin{document}

\title{Strong decays of charmonium-like state Y(4230)}

\author{Gurjav Ganbold}
\email{ganbold@theor.jinr.ru}
\affiliation{Bogoliubov Laboratory of Theoretical Physics,
Joint Institute for Nuclear Research, 141980 Dubna, Russia}
\affiliation{Institute of Physics and Technology, Mongolian Academy
of Sciences, 13330 Ulaanbaatar, Mongolia}

\author{M. A. Ivanov}
\email{ivanovm@theor.jinr.ru}
\affiliation{Bogoliubov Laboratory of Theoretical Physics,
Joint Institute for Nuclear Research, 141980 Dubna, Russia}

\begin{abstract}
Strong decays of the charmonium-like state $Y(4320)$ have been studied in the
framework of the covariant confined quark model. The resonance $Y$ has been
interpreted  as a four-quark state of the molecular  type.  We evaluate the
hidden-charm decay width of $Y$ to a vector and a scalar, with the latter
decaying subsequently to a pair of charged pseudoscalar  states. The strong
decay modes $Y \!\to\! \pi^{+}\pi^{-}J/\Psi$ and $Y \!\to\! K^{+}K^{-} J/\Psi$
have been studied by involving the scalar resonance state $f_0(980)$. We have
calculated the fractal widths of the related strong decays and the branching  ratio
${\cal B}(Y \!\to\! K^{+}K^{-} J/\Psi)$ / ${\cal B}(Y \!\to\! \pi^{+}\pi^{-}J/\Psi)$, 
recently determined by the BESIII  collaboration. The estimated branching ratio
and calculated fractal strong decay widths are in good agreement with the latest
experimental data, favoring the molecular representation of the $Y(4320)$ state.

\pacs{13.20.Gd,13.25.Gv,14.40.Rt,14.65.Fy, 14.40.Lb, 13.40.Hq,
 12.39.-x, 12.38.Aw, 14.65.Dw}

\keywords{relativistic quark model, confinement, exotic states,
charmonium, tetraquarks, decay widths}

\end{abstract}

\maketitle

\section{Introduction}
\label{sec:intro}

During the past years, several unusual states were observed in the course
of experimentally establishing the heavy meson spectrum. They possess
a common feature - a minimal constituent quark-antiquark model does not
work for these states.  One of these unusual states is the $Y(4230)$
(also known as $\psi(4260)$ and $Y(4260)$) and this resonance still needs
solid experimental study.
Historically, the data analysis of the mass spectrum  of $\pi^+\pi^-\Jpsi$ in
production $e^+e^-\to\gamma_{\rm ISR}\, \pi^+\pi^-\Jpsi$ by the BaBar
Collaboration observed a broad resonance around $4.26$~GeV
\cite{Aubert:2005rm}. The spin-parity $J^{PC}=1^{--}$ indicates a
possible charmonium state, but its mass did not match any known mass of
charmonium states. Along with the strong coupling to $\pi^+\pi^-\Jpsi$,
no evidence was found for coupling to any open charm decay modes.

Earlier studies suggested that the $Y(4230)$ state is not a conventional state
of charmonium~\cite{Brambilla:2010cs}. Then, $Y$ has been interpreted as
a charmonium-hybrid state in Refs.~\cite{Zhu:2005hp,Kou:2005gt,Close:2005iz}.

An interpretation of $Y(4260)$ as the first orbital excitation of
diquark-antidiquark state $([cs][\bar c \bar s])$ was proposed in Ref.
~\cite{Maiani:2005pe} by assuming a dominant decay to $D_s\bar D_s$.

Another way to look at $Y(4260)$ is as a $\chi_{c1}-\rho^0$ molecule
was developed in ~\cite{Liu:2005ay}. Then, the $Y(4260)\to\pi^+\pi^-\Jpsi$
decay width should be  larger than $Y(4260)\to D\bar D$ that has not been
observed.
The absence of the $Y(4260)$ in the decays with open charm can be clearly
explained by an assumption considering the $Y$ as a $D\bar D_1$ molecular
state~\cite{Wang:2013cya}.
In the framework of the molecular-type interpretation,
some hidden-charm and charmed pair decay channels via intermediate
$DD_1$ meson loops have been investigated ~\cite{Li:2013yla}. Considering
$DD_1$, a weakly bound state, a two-body decay $Y(4260)\to Z_c(3900) + \pi$
has been studied. Furthermore, the decay mode $Y(4260) \to \Jpsi +\pi^+\pi^- $
was also calculated \cite{Dong:2013kta}.

The cross sections of $e^+e^- \rightarrow K^+K^-J/\psi$ have more recently
been  measured by  using 15.6 fb$^{-1}$ gathered with the BESIII detector
\cite{Ablikim:2022}. The mass and width of this structure are measured to be
($4225.3~\pm~2.3~\pm~21.5$) MeV and ($72.9 \pm 6.1 \pm 30.8$)~MeV,
respectively. They align with the standards set forth for $Y(4230)$.

Furthermore, the ratio of the branching fractions of $Y(4230)$ decaying into
$K\bar{K}J/\psi$ and $\pi \pi J/\psi$ modes is determined \cite{Ablikim:2022}
to be
$$
0.02 < \frac{\mathcal{B}(Y(4230) \to K^+K^-J/\psi)}
{\mathcal{B}(Y(4230) \to \pi^+\pi^-J/\psi)} < 0.26 \,,
$$
where ranges are determined from the eight combinations of multiple solutions
in the two measurements and the large ambiguity is primarily due to the multiple
solutions in the $\pi^+\pi^-\Jpsi$ measurement. However, the difficulties in reaching
a judgment before finding the physics solution in the measurements has been
highlighted, particularly for the $\pi^+\pi^-\Jpsi$ mode.

In the present paper we investigate strong decays of the $Y(4320)$ resonance
in the framework of the covariant confined quark model (CCQM)~\cite{Branz:2009cd}
proposed and developed by us last decade. The model implements effective
quark confinement with convolutions of local quark propagators and vertex functions
accompanied by an infrared cutoff of the scale integration that prevents any
singularities in matrix elements. We have applied the CCQM to multiquark meson
structure, form factors and angular decay characteristics of light and heavy  baryons.
Possible new physics effects in the exclusive decays of $\bar{B}^0$  and transitions
of $\Lambda_b$ have been studied in some extension of the Standard Model by
taking into account right-handed vector (axial), left- and right-handed (pseudo)scalar,
and tensor current contributions.
Inspired by recent measurements we have studied the radiative decays of
charmonium  states below the $D {\bar D}$ threshold by introducing only one
adjustable parameter common for the six charmonium states. The obtained
results were in good agreement with the latest data,  we also predicted a more
narrow full width for the $h_c$ charmonium~\cite{Ganbold:2021}.
In our papers devoted to description of the multi-quark states
Refs.\cite{Dubnicka:2011mm}, first, we have treated the $X(3872)$ meson
as a diquark-antidiquark bound state. In ~\cite{Goerke:2016hxf}  we have
employed a molecular-type four-quark current to describe the decays of
the $Z_c (3900)$ state and shown that a molecular-type current  can give
the decay width values in accordance with the experiment.  By using
molecular-type four-quark currents for the recently observed resonances
$Z_b(10610)$ and $Z_b(10650)$, we have calculated in
Ref.~\cite{Goerke:2017svb} their two-body decay rates into
a bottomonium state plus a light meson as well as into B-meson pairs.
A brief sketch of our findings may be found in Ref.~\cite{Ivanov:2018ayq}.
We have investigated two interpolating currents for the $Y(4260)$ resonance:
the molecular-type current and the  tetraquark one~\cite{Dubnicka:2020}.
We demonstrated that the mode  $Y\to Z^+_c + \pi^-$  was enhanced in both
approaches when compared to open-charm modes.

The paper is organized as follows.
A brief description of the CCQM and the general formalism for describing
$Y(4260)$ as four quark molecular state are give in Sec.~II.
In Sec.~III we calculate the related amplitude and width of the strong decay
$Y\to f_0 + \Jpsi$. Sec.~IV is dedicated to the investigation of  two decays:
$f_0 \!\to\! \pi^{+} \!+\! \pi^{-}$ and $f_0 \!\to\! K^{+} \!+\! K^{-}$ with related
formulas for the Lorentz form factors, matrix elements, phase volumes and
fractional decay widths. In the next, Sec.~V, we study
the strong decays $Y(4230) \to  K^{+}  K^{-}  \Jpsi $ and
$Y(4230) \to \pi^{+} \pi^{-} \Jpsi $ occuring via intermediate scalar resonance
$f_0(980)$ in different ways: by using the narrow-width approximation, its
improved modification  and full integration over kinematically allowed
resonance momentum.  In Sec.~VI we discuss the obtained results by
comparing them with latest experimental data.

\section{Approach}
\label{sec:approach}

The covariant confining quark model (CCQM) ~\cite{Branz:2009cd} represents an
effective quantum field treatment of hadronic effects and derived from Lorentz
invariant non-local Lagrangian in which a hadron is coupled to its constituent quarks.
Hereby, hadrons are characterized by size parameters $\Lambda_H$ dictating the
strength of the quark-hadron coupling. This mechanism is done by using so-called
compositeness condition ~\cite{Salam:1962ap, Weinberg:1962hj} requiring the
wave-function renormalization constant of the hadron to be zero $Z_H=0$. This
condition reduces the number of free parameters (i.e. couplings) and guarantues
a correct description of bound states as dressed (with no overlap with bare states).
The hadron-quark interaction vertices are given by a Gaussian-type functions.
An analytic built-in confinement, based on a cutoff in the integration space of
Schwinger parameters of quark propagators, removes all divergencies and  the model
can be used for description of arbitrary heavy hadrons.

Below we use the CCQM to investigate three modes of the strong $Y(4260)$-decay
with hidden-charm combinations as follows:
$Y\to \Jpsi \, f_0 $, $Y\to \Jpsi \, K^{+} \, K^{-}$ and  $Y\to \Jpsi \, \pi^{+} \, \pi^{-}$,
shown in Figs. 2-5.

According to the CCQM, the effective interaction Lagrangian describing the
coupling of the meson $Y(4230)$ to its constituent quarks can be written in
the form:
\be
{\cal L}_{\rm int} = g_{Y}\, Y_{\mu}(x)\cdot J^\mu_{Y}(x) + \text{H.c.}
\label{eq:lagran}
\en

Since the neutral state $Y(4230)$  with the quantum numbers
$I^G(J^{PC}) = 0^{-}(1^{--})$ cannot be successfully described within the framework
of any quark-antiquark bound state,  we introduce the interpolating four-quark
molecular-type current as follows:
\be
J_Y^\mu = \frac{1}{\sqrt{2}}
\left\{ (\bar q \gamma_5 c) (\bar c\gamma^\mu \gamma_5 q)
      + (\gamma_5\leftrightarrow \gamma^\mu\gamma_5) \right\}
\label{eq:Y4cur}.
\en

The corresponding nonlocal generalization of the four-quark current
within the CCQM reads
\bea
\label{eq:Y4nonloc}
J_{Y}^\mu (x) &=& \int\! dx_1\ldots \int\! dx_4 \delta
\left( x-\sum\limits_{i=1}^4 w_i x_i \right)
\Phi_{\,Y}\Big(\sum\limits_{i<j} (x_i-x_j)^2 \Big)
J_{Y_{non}}^\mu (x_1,\ldots,x_4),  \\
J_{Y_{non}}^\mu &=&  \tfrac{1}{\sqrt{2}} \Big\{
 (\bar q(x_3) \gamma_5 c(x_1))\cdot (\bar c(x_2) \gamma^\mu\gamma_5 q(x_4) )
+ (\gamma_5\leftrightarrow \gamma^\mu\gamma_5)\Big\},
\qquad (q=u,d), \nonumber
\ena
where the reduced quark masses $w_i=m_i/\sum_{j=1}^4 m_j$ are specified
for $m_1=m_2 = m_c$ and $m_3=m_4 = m_q$. Hereby, we neglect the
isospin violation in the $u-d$ sector, i.e. $m_u=m_d=m_q$. The numbering of
the coordinates $x_i$ is chosen such that one has a convenient arrangement of
vertices and propagators in the Feynman diagrams to be calculated.

The translationally invariant four-quark vertex function reads:
\be
\Phi_Y \Big(\sum\limits_{i<j} (x_i-x_j)^2 \Big) =
\prod\limits_{i=1}^3\int\!\!\frac{d^4 q_i}{(2\pi)^4}\, e^{ - i q_i ( x_i - x_4 ) }
\widetilde\Phi_Y (- Q^2),  \qquad  Q^2=\frac12\sum\limits_{i\le j}q_iq_j .
\label{eq:Y4vertex}
\en

The Fourier transform of the translationally invariant vertex function in momentum
space is required to fall off in the Euclidean region in order to provide the ultraviolet
convergence of the loop integrals.  We use a simple Gaussian form as follows:
\be
\widetilde\Phi_{Y}(- Q^2) = \exp(Q^2 / \Lambda^2_{Y} ) \,,
\label{eq:Y4vertsize}
\en
where $\Lambda_{Y} >0 $ is an adjustable hadron size-related parameter of the
CCQM. In fact, any choice for  $\widetilde\Phi_{Y}$ is appropriate as long as it
falls off sufficiently fast in the ultraviolet region to render the corresponding
Feynman diagrams ultraviolet finite.

The renormalized coupling constant $g_{Y}$ in Eq.~(\ref{eq:lagran}) is
strictly determined by the {\it compositeness condition}
(see Ref. \cite{Branz:2009cd} for details) as follows:
\be
\label{eq:renorm}
Z_{Y} = 1 - g^2_{Y}\,\frac{d}{dp^2} \widetilde\Pi_{Y}(p^2) = 0,
\qquad p^2 = M^2_Y,
\en
where $\Pi_{Y}(p^2)$ is the diagonal (scalar) part of the vector-meson
mass operator defined in momentum space
\be
\widetilde\Pi_{Y}(p^2) = \frac{1}{3} \left(g_{\mu\nu}
- \frac{p_\mu p_\nu}{p^2}\right)\widetilde\Pi^{\mu\nu}_{Y}(p).
\label{eq:self0}
\en
Therefore, any bare states are removed totally from consideration, the mass
and  wave function of the hadron are renormalized, and  the physical state is
dressed.

\begin{figure}[ht]
\begin{center}
\epsfig{figure=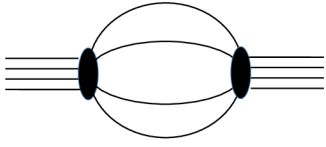,scale=4.0}
\caption{Feynman diagram for $Y(4230)$ mass operator}
\label{fig:FIG1}
\end{center}
\end{figure}

The Fourier transform of the mass operator for the four-quark state
$Y$ is written as
\bea
\widetilde\Pi_{Y}^{\mu\nu}(p) &=&
\frac{N_c^2}{2} \,\prod\limits_{i=1}^3\int\!\!\frac{d^4k_i}{(2\pi)^4i}\,
\widetilde\Phi_{Y}^2\left(
-(k_1^2 + k_2^2 + k_3^2 - k_1 k_2  - k_1 k_3 + k_2 k_3)/2
\right)
\nn
&\times&
\Big\{
\Tr\left[ S_3(\hat k_3 + w_3 \hat p)\gamma_5
             S_1(\hat k_1-  w_1 \hat p) \gamma_5  \right]
\\ \nonumber
&&
\times \Tr\left[ S_2(\hat k_2 + w_2 \hat p)\gamma^\mu \gamma_5
             S_4(- \hat k_1 + \hat k_2 + \hat k_3 - w_4 \hat p) \gamma^\nu \gamma_5\right]
\\ \nonumber
&&
+ \,
\Tr\left[ S_3(\hat k_3 + w_3 \hat p)\gamma_5
             S_1(\hat k_1-  w_1 \hat p) \gamma^\nu \gamma_5  \right]
\\ \nonumber
&&
\times \Tr\left[ S_2(\hat k_2 + w_2 \hat p)\gamma^\mu \gamma_5
             S_4(- \hat k_1 + \hat k_2 + \hat k_3 - w_4 \hat p) \gamma_5\right]
\Big\}                \,,              \quad N_c = 3 \,.  \nonumber
\label{eq:self}
\ena

For the quark propagator we use the Fock-Schwinger representation:
\be
\widetilde{S}_j(\hat{k}) =  \left( m_j + \hat{k} \right) \intop_{0}^{\infty}
d\alpha\: \exp\big(-\alpha \left(m_j^2 - k^2 \right) \big) \,.
\label{green}
\en

Matrix elements of hadronic processes described by using Feynman
quark-loop diagrams may be represented in terms of convolutions of vertex
functions and quark propagators as follows:
\begin{equation}
\label{integ0}
\Pi^{0} \!\!= N_c \!\!\int_0^\infty\! \!\! dt \, t^{n-1} \!\!\int_0^1\! \!\! d^n
\alpha \, \delta\Big( 1 \!\!-\!\! \sum_{i=1}^n \alpha_i \Big)
f(t\alpha_1,\ldots,t\alpha_n).
\end{equation}
Hereby,  possible branch points connected with the creation of free quarks
may appear. These threshold singularities can be removed by introducing
a universal infrared cutoff parameter, $\lambda$ as follows:
$\!\!\int_0^\infty \!\! dt  \ldots   \to \int_0^{1/\lambda^2} \!\! dt  \ldots $.

\section{Strong decay $Y(4230) \to \Jpsi + f_0(980)$}
\label{sec:YVS}

Consider the strong decay $Y(4230) \to \Jpsi + f_0(980)$ represented
in the leading order by the Feynman diagram in Fig.1.

\begin{figure}[ht]
\begin{center}
\epsfig{figure=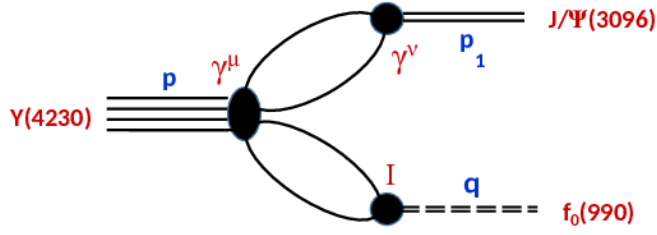,scale=8.0}
\caption{Leading-order Feynman diagram for decay $Y \to \Jpsi + f_0$.}
\label{fig:FIG2}
\end{center}
\end{figure}

The matrix element of the two-body decay  of $Y$ into a vector ($V=\Jpsi(3096)$)
and a scalar ($S=f_0(980)$)  reads
\be
\label{eq:MatYVS}
{\cal M}_{YVS} =
{\cal M}
\left( Y(p,\epsilon^{\mu}_p) \to \Jpsi(p_1,\epsilon^{\nu}_{p_1})+ f_0(q)\right)
\doteq \,g_{Y} \, g_{\Jpsi} \, g_{f_0} \, \epsilon^{\mu}_p \, \epsilon^{\ast\,\nu}_{p_1}
\left( d_A \, p_1^\mu p^\nu - d_B \,g^{\mu\nu} (p\,p_1) \right) \,.
\en

The Lorentz form factors $d_A$ and $d_B$ in the LO are defined as follows:
\bea
&&  d_A \, p_1^\mu p^\nu - d_B \,g^{\mu\nu}
      =       \nn[2ex]
&& \frac{N_c^2}{2}
\int\!\!\frac{d^4k_1}{(2\pi)^4i}\,\int\!\!\frac{d^4k_2}{(2\pi)^4i}\,
\widetilde\Phi_{Y}\left(-Q^2\right)
\widetilde\Phi_{\Jpsi}\left(-\,(\ell_1 + \ell_2)^2/4\right)
\widetilde\Phi_{f_0}\left(-\,(\ell_3 + \ell_4)^2/4\right)
\nn
&\times&
\Big\{
\Tr\left[ \gamma_5   S_1(\ell_1)\gamma^\nu S_2(\ell_2)
              \gamma^\mu\gamma_5 S_4(\ell_4) I S_3(\ell_3) \right]
+\Tr\left[\gamma^\mu \gamma_5   S_1(\ell_1)\gamma^\nu S_2(\ell_2)
              \gamma_5 S_4(\ell_4) I S_3(\ell_3) \right]
              \Big\} \,, \nn
&&
Q^2 = \left[ (\ell_1+p w_1)^2+ (\ell_2-p w_2)^2+(\ell_3+p w_3)^2
       +(\ell_4-p w_4)^2\right] /2, \nn
&&
\ell_1 = k_1 - p w_1, \quad \ell_2 = k_1 - p w_1 - p_1, \quad
\ell_3 = k_2 - p w_4 - q, \quad \ell_4 = k_2 - p w_4\,,
\label{eq:dAdB}
\ena
where $\widetilde\Phi_{\Jpsi}(- k^2) = \exp(k^2 / \Lambda^2_{\Jpsi})$ and
$\widetilde\Phi_{f_0}(- k^2) = \exp(k^2 / \Lambda^2_{f_0})$ are vertex
functions with adjustable hadron size-related parameters
$\Lambda_{\Jpsi}$ and $\Lambda_{f_0}$.

The fractional strong decay width $\Gamma (Y\!\! \to\!\! \Jpsi \! + \! f_0)$ is
calculated as follows:
\bea
\label{eq:GamYVS}
\Gamma_{YVS}
\!\!\!\!&&
= \frac{g^2_{Y}g^2_{\Jpsi}g^2_{f_0} }{(2s + 1)\, 2 \, M_Y}
\, (p\, p_1)^2
\left\{ (\xi\!-\!2\!+\!\xi^{-1}) \, d_A^2  \!+\! 2\, (1\!-\!\xi) \,  d_A d_B
  \!+\! (2\!+\!\xi) \,  d_B^2 \right\} \cdot \Omega_{YVS} \,, \\
&& (p\, p_1)^2 = (p^2 + p_1^2 - q^2)/2\,, \qquad \xi
\doteq (p\, p_1)^2 /(p^2\,p_1^2) \,,  \nonumber
\ena
where $s=1$ is the spin of unpolarized particle Y(4230) and the phase
volume of the decay is introduced as follows:
\be
\label{eq:PhasYVS}
\Omega_{YVS} =
\Omega_{Y \to \Jpsi + f_0} (M_{Y}^2, M_{\Jpsi}^2, M_{f_0}^2) =
\frac{1}{8\pi M_Y^2} \left[ \lambda(M_{Y}^2, M_{\Jpsi}^2, M_{f_0}^2) \right]^{1/2}
\en
with the K{\"{a}}llen function $\lambda(p^2, p_1^2, q^2)$.

\section{Strong decay of $f_0(980)$ into two pseudoscalar mesons}
\label{sec:SPP}

Let us consider a strong decay of a scalar particle
$f_0(q) \to P^{+}(q_1) + P^{-}(q_2)$ into a pair of
charged pseudoscalar mesons - either $\pi^+\pi^-$, or $K^+K^-$. Further,
we use notations $P^{+}=\{\pi^+, \, K^+ \}$ and $P^{-}=\{\pi^-, \, K^- \}$.

The matrix element of the decay takes the following form:
\be
{\cal M}_{SPP} = g_{f_0} \, g_{P^{+}} \, g_{P^{-}} \cdot  d_C \,.
\label{eq:MatSPP}
\en

The corresponding Lorentz form factor of the decay in the LO reads
\bea
d_C &=& \frac{N_c}{2} \int\!\!\frac{d^4k}{(2\pi)^4i}\,
\widetilde\Phi_{f_0}\left( -k^2 \right)
\widetilde\Phi_{P^{+}}\left(-\,(\ell_2 + \ell_3 + q_1 (w_s-w_q) )^2/4\right) \nn
&\times&
\widetilde\Phi_{P^{-}}\left(-\,(\ell_1 + \ell_3 - q_2 (w_s-w_q))^2/4\right)
\Tr\left[ \gamma_5   S_1(\ell_1) I  S_2(\ell_2) \gamma_5 S_3(\ell_3) \right]  \,,
\label{eq:dC}
\ena
where $\ell_1 = k - q/2, \quad \ell_2 = k+q/2, \quad \ell_3 = k - q_1/2 +q_2/2$
and $\widetilde\Phi_{f_0}(-k^2) = \exp(k^2/\Lambda_{f_0})$,
$\widetilde\Phi_{P^{+}}(-k^2) = \exp(k^2/\Lambda_{P^{+}})$,
$\widetilde\Phi_{P^{-}}(-k^2) = \exp(k^2/\Lambda_{P^{-}})$ with corresponding
'size' parameters $\Lambda_{f_0}$,  $\Lambda_{P^{+}}$ and $\Lambda_{P^{-}}$.

\begin{figure}[ht]
\begin{center}
\epsfig{figure=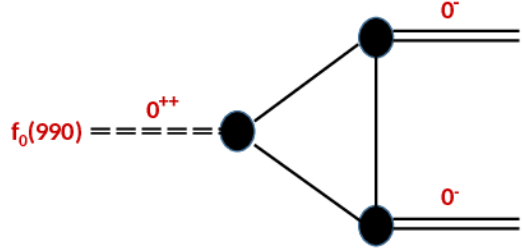,scale=8.0}
\caption{Leading-order Feynman diagram for strong decay of $f_0$
into two pseudoscalar mesons.}
\label{fig:FIG3}
\end{center}
\end{figure}

The fractional strong decay widths is calculated as follows:
\be
\Gamma_{SPP} =
\Gamma (f_0 \!\to\! P^{+} \!+\! P^{-}) =
\frac{g_{f_0}^2 \, g_{P^{+}}^2 \, g_{P^{-}}^2}{2 \cdot 2 \, M_S} \cdot d_C^2
 \cdot \Omega_{SPP} \,,
\label{eq:GamSPP}
\en
where the phase volume of the decay is introduced as follows:
\be
\Omega_{SPP} =
\Omega_{f_0 \!\to\! P^{+} \!+\! P^{-}} (M_{f_0}^2, M_{P^{+}}^2, M_{P^{-}}^2)
= \frac{1}{8\pi M_{f_0}^2}
\left[ \lambda(M_{f_0}^2, M_{P^{+}}^2, M_{P^{-}}^2) \right]^{1/2}
\label{eq:PhasSPP}
\en
with the K{\"{a}}llen function $\lambda(q^2, q_1^2, q_2^2)$.

\newpage
\section{Strong decays of $Y(4230) \to  K^{+}  K^{-}  \Jpsi $ and
                                        $Y(4230) \to \pi^{+} \pi^{-} \Jpsi $}
\label{sec:YVPP}

We consider the following sequential two-body decays: first the four-quark
charmonium-like state decays into the vector charmonium $\Jpsi(3096)$ and
the scalar resonance $f_0(980)$, then,  $f_0$ decays into two charged
pseudoscalar mesons - either $\pi^+\pi^-$, or $K^+K^-$.

\begin{figure}[ht]
\begin{center}
\epsfig{figure=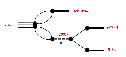,scale=8.0}
\caption{Leading-order Feynman diagram for decays $Y \to \pi^{+} \pi^{-} \Jpsi $
and $Y \to  K^{+}  K^{-}  \Jpsi$.}
\label{fig:FIG4}
\end{center}
\end{figure}

The corresponding decay width is given by calculating an integral:
\bea
&&
\Gamma (Y \!\to\! P^{+} P^{-} \Jpsi) = \frac{1}{24 M_Y }
\nn
&\times& \!\!\!\!
\!\!\!\!\int\limits_{q^2_{min(P^+P^-)}}^{q^2_{max}} \!\!\!\!\!\! dq^2 \,
\left[ {\cal M}_{YVS}(q^2) \right]^2 \, \Omega_{YVS}(q^2) \, BW(q^2)
\, \frac{1}{\sqrt{q^2}} \, \left[ {\cal M}_{SPP}(q^2) \right]^2 \, \Omega_{SPP}(q^2) \,
\label{eq:GamYVPP}
\ena
where the integration regions are bound by the corresponding kinematic
limits.

The Breit-Wigner resonance distribution function is introduced as follows:
\be
BW(q^2) = \frac{1}{\pi} \frac{M_S}{(q^2-M_S^2)^2+(M_S \Gamma_S)^2} \,,
\qquad
\int\limits_{-\infty}^{+\infty} \!\! dq^2 \, BW(q^2) = \frac{1}{\Gamma_S} \,.
\label{eq:BW}
\en

Below we calculate the branching ratio of strong decays with hidden charms,
${\cal B}(Y \!\to\! K^{+}K^{-} \Jpsi)$ / ${\cal B}(Y \!\to\! \pi^{+}\pi^{-}\Jpsi)$,
via resonance state $f_0(980)$ in different ways, namely, by using the
narrow-width approximation (NWA), its phase-space improved modification
(PSINWA) and direct integration over kinematically allowed momentum of
the resonance.

\subsection{Narrow Width Approximation (NWA) }
\label{sec:NWA}

The integration over $q^2$ in Eq.(\ref{eq:GamYVPP}) may be directly
calculated by numerical means.

However, taking into account the latest data
($\Gamma_{f_0}=10 \div 100$ MeV and $M_{f_0}=990$ MeV)  one can
try to apply the NWA scheme by using an approximation:
\be
BW(q^2) \sim \frac{1}{\Gamma_S} \, \delta(q^2-M_S^2)
\en
which is effective for $\Gamma_S/M_S \ll 1$   (see, e.g. \cite{Uhlemann:2008}).

Then, one obtains rough estimates for the decays width under consideration
as follows:
\be
\Gamma_{NWA} (Y \!\to\! P^{+} P^{-}\Jpsi) =
\Gamma_{YVS} \cdot {\cal B}(S \!\to\! P^{+} P^{-}) \,, \qquad
{\cal B}(S \!\to\! P^{+} P^{-}) \doteq
\frac{\Gamma (S \!\to\! P^{+} P^{-})}{\Gamma_S} \,.
\label{eq:GamNWA}
\en

Accordingly, the branching ratio in the NWA reads:
\be
\left[ \frac{{\cal B}(Y \!\to\! K^{+} K^{-}\Jpsi)}
{{\cal B}(Y \!\to\! \pi^{+} \pi^{-}\Jpsi)} \right]_{NWA}
= \frac{\Gamma(S \!\to\! K^{+} K^{-})}{\Gamma(S \!\to\! \pi^{+} \pi^{-}) } \,,
\quad M_s = M_{f_0}=990 \text{MeV} \,.
\label{eq:BRNWA}
\en

However, it has been observed that the NWA can be unreliable in relevant
circumstances, namely with decays where a daughter mass approaches the
parent mass \cite{Uhlemann:2008}, that exactly occurs for the decay
$f_0 \!\to\! K^{+} K^{-}$. Second, the uncertainty of the NWA is commonly
estimated as nearly  $\sim 3\,\Gamma_S / M_S$, i.e. of order $\sim 30 \%$
in the case of $f_0(980)$.

\subsection{Phase-space Improved Narrow Width Approximation (PSINWA) }
\label{sec:PSINWA}

For a more accurate calculation, using a modified NWA prescription may be more
efficient. According to the phase-space improved narrow-width approximation
(PSINWA) in \cite{Uhlemann:2008}, one should substitute the mass of the
resonance $M_S$  with an effective mass  $M_{eff}$ instead in the NWA formula.
Hereby, in analogy to $M_S^2$ being the maximum position of the Breit-Wigner,
$M_{eff}^2$ is given by the position of the maximum of the production of
Breit-Wigner and the phase-space factors of the related decay processes.

\begin{figure}[ht]
\begin{center}
\epsfig{figure=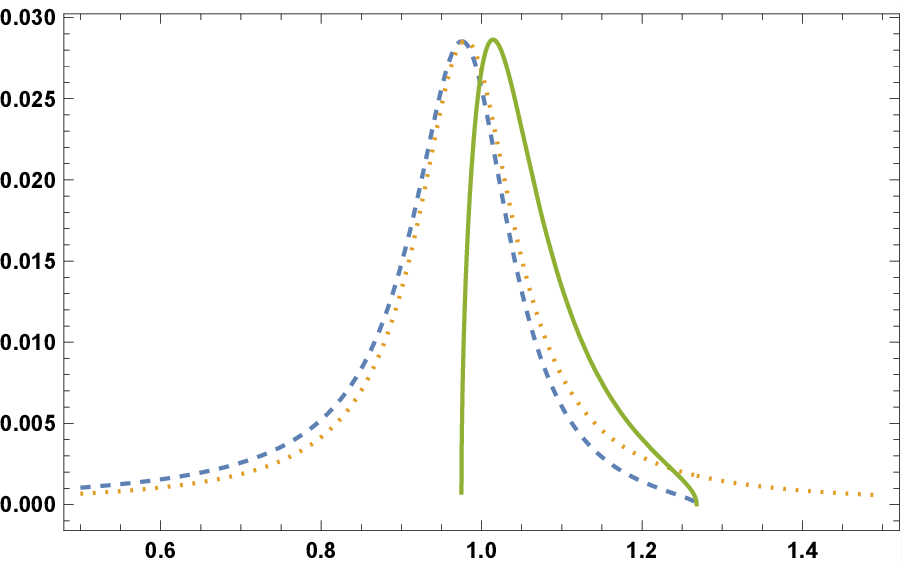,scale=.55}
\epsfig{figure=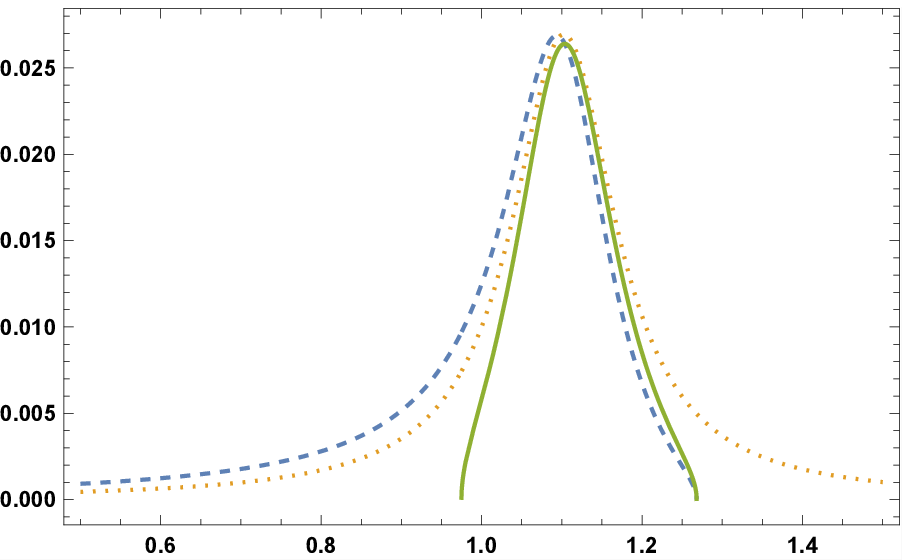,scale=.55}
\epsfig{figure=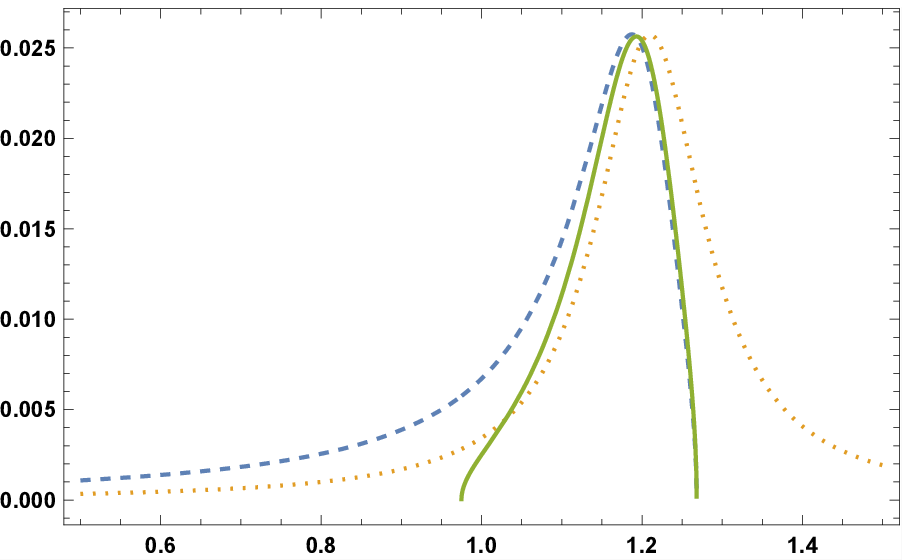,scale=.55}
\caption{Shape of the pure Breit-Wigner curve (yellow, dotted) vs the
production  of Breit-Wigner by the phase-space factors of strong decay
$Y \to  K^{+}  K^{-}  \Jpsi$  (green, solid) and  $Y \to  \pi^{+}  \pi^{-}  \Jpsi$
(blue, dashed) for different resonant effective masses
$M_{eff}=$  990, 1050 and 1100 MeV.
All curves are reduced to the same height for better visibility.
}
\label{fig:FIG5}
\end{center}
\end{figure}

By varying the effective mass, we reveal (see in Fig.5) that the maximum of
the production of Breit-Wigner and the phase-space factors of strong decay
$Y \to  K^{+}  K^{-}  \Jpsi$ is occured at $M_{eff}=1050$ MeV.

We calculate an improved  branching ratio in the PSINWA approach as follows:
\be
\left[ \frac{{\cal B}(Y \!\to\! K^{+} K^{-}\Jpsi)}
               {{\cal B}(Y \!\to\! \pi^{+} \pi^{-}\Jpsi)} \right]_{PSINWA}
= \frac{\Gamma(S \!\to\! K^{+} K^{-})}{ \Gamma(S \!\to\! \pi^{+} \pi^{-}) }
|_{M_{S} = M_{eff}=1050 \text{MeV}} \,.
\label{eq:BRPSI}
\en
Corresponding numerical result is given in Tab. I.

\subsection{Full Integration over resonance momentum}
\label{sec:integ}

By directly integrating the expression in Eq.~(\ref{eq:GamYVPP}), we can
accurately calculate the branching ratio and obtain the following estimate:
\be
\left[ \frac{{\cal B}(Y \!\to\! K^{+} K^{-}\Jpsi)}
               {{\cal B}(Y \!\to\! \pi^{+} \pi^{-}\Jpsi)} \right]_{integ}
= \frac{\Gamma (Y \!\to\! K^{+} K^{-} \Jpsi)}{\Gamma (Y \!\to\! \pi^{+} \pi^{-} \Jpsi)} \,.
\label{eq:BRint}
\en
The related kinematic limits of the decays constrain the integration regions
as follows:
\be
                q^2_{max} = (M_Y - M_{\Jpsi})^2 \,,      \qquad
q^2_{min(\pi^+\pi^-)} =  4 M_{\pi}^2 \,,   \qquad
   q^2_{min(K^+K^-)} =  4 M_{K}^2 \,.
\en

\section{Numerical results and Conclusion }
\label{sec:result}

The correct determination of the renormalized couplings $g_{H}$
($H=\{Y, \Jpsi, f_0, K, \pi \}$) of the participating hadrons is the first step
in calculating physical observable decay widths. According to the CCQM,
the renormalized couplings  are strictly fixed by the compositeness
requirements expressed in Eq.~(\ref{eq:renorm}) and play an important role
by excluding the constituent degrees of freedom from the space of physical
states.

Using Eqs.~(\ref{eq:GamYVS}), (\ref{eq:GamSPP}), (\ref{eq:GamYVPP}),
(\ref{eq:BRNWA}), (\ref{eq:BRPSI}) and (\ref{eq:BRint}) we can determine
the partial widths of the strong decays of the $Y$ resonance and their
branching ratios after computing the necessary renormalization couplings,
$g_H$.

The model parameters of the CCQM are determined by minimizing $\chi^2$
in a fit to the available experimental data and some lattice results. The fitted
parameters may vary around their central value by about $\pm 10 \%$, and the
errors of our calculations do not exceed  also ten percent. The updated
central values may be found, e.g., in ~\cite{Dubnicka:2020,Ganbold:2021}.
In particular, we have set the charm quark mass and the $\Jpsi$-charmonium
size parameter as $m_c=1.80$~GeV and $\Lambda_{\Jpsi} \simeq 1.55$~GeV,
respectively, in our latest study on the low-lying charmonium states
~\cite{Ganbold:2021}.

Below, we use the following basic parameter values (in GeV):
\begin{equation}
\begin{tabular}{ c c c c c c }
\quad $\lambda$ \quad & \quad $m_{u/d}$ \quad & \quad $m_s$
                             \quad & \quad $m_c$ \quad & \quad $\Lambda_{f_0}$
                             \quad & \quad $\Lambda_{\Jpsi}$
\quad
\\\hline
\quad 0.181 \quad & \quad 0.264 \quad & \quad 0.390
                     \quad & \quad  1.80 \quad & \quad 1.25
                     \quad & \quad 1.55
\quad \,.
\end{tabular}
\label{basicpara}
\end{equation}

We can adjust the size parameters $\Lambda_{\pi}$, $\Lambda_{K}$,
$\Lambda_{\Jpsi}$ and $\Lambda_{Y}$ to fit the  latest data
 ~\cite{PDG2022,Ablikim:2022}:
\be
 \Gamma_{Y} = 49 \pm 8 \text{MeV} \,, \quad
 \Gamma_{f_0} = 10 \div 100 \text{MeV} \,,  \quad
\left[ \frac{{\cal B}(Y \!\to\! K^{+} K^{-}\Jpsi)}
{{\cal B}(Y \!\to\! \pi^{+} \pi^{-}\Jpsi)} \right] = 0.02 \div 0.26 \,.
\en

1) The experimental data for the $\Gamma_{f_0}=10\div 100 \text{MeV}$ is
spread too wide, and it can be narrowed by taking the following into account.
According to \cite{PDG2022} the peak width in $\pi\pi$ is about $50$ MeV, but
decay width can be much larger.  On the other hand, the branching ratio
$\Gamma(f_0\to \pi^+ \pi^-)/[\Gamma(f_0\to \pi^+ \pi^-)+\Gamma(f_0\to K^+ K^-)]$
is about 0.75$\pm$0.13 (see, e.g. \cite{PDG2022}).  By assuming that these
results are accurate, we can significantly reduce the interval:
$60 <  \Gamma_{f_0} < 100 \text{MeV}$. Further, we will exploit  the following
average value $ \Gamma_{f_0} = 80 \text{MeV}$.

\begin{figure}[htb]
\begin{center}
\epsfig{figure=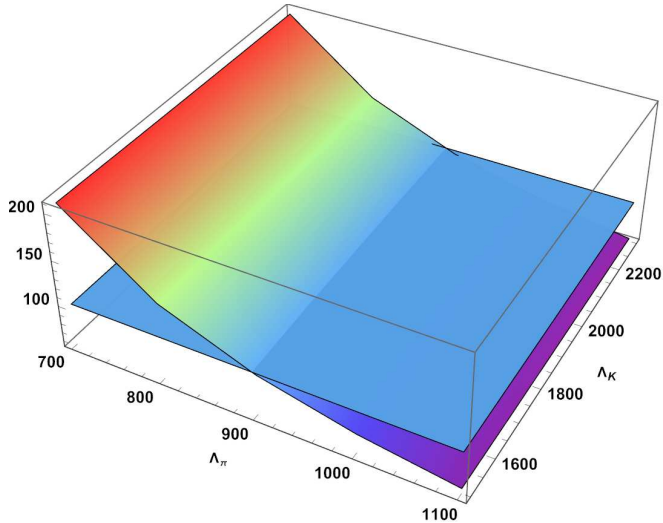,scale=0.95}
\caption{Dependence of the sum
$\Gamma(f_0\!\to\! K^+ K^-) \!+\! \Gamma(f_0\!\to\! \pi^+ \pi^-)$
on size parameters $\Lambda_\pi$ and $\Lambda_K$.
}
\label{fig:FIG6}
\end{center}
\end{figure}

2) The size parameters $\Lambda_{\pi}$ and $\Lambda_{K}$ significantly affect
the widths of  sub-processes $f_0 \to P^+ P^-$. Particularly, the dependence of
the sum $\Gamma(f_0\to K^+ K^-) + \Gamma(f_0\to \pi^+ \pi^-)$ expressed in MeV
on $\Lambda_\pi$ and $\Lambda_K$ is shown in Fig. 6. We reveal that the
constraint $\Gamma_{f_0} < 100$ MeV is achieved for the size parameters
$\Lambda_{\pi} > 900$ MeV and $\Lambda_{K} \sim 1700 \div 2200$ MeV.

3) The value of $\Lambda_{Y}$ mostly affects the main decay width values of
$Y\to f_0 \Jpsi$ and $Y\to \Jpsi P^+ P^-$.  Corresponding dependence is shown
in Fig.7. One can see that the requirements
$\Gamma(Y\to f_0 \Jpsi) < \Gamma_{Y} $ and
$\Gamma(Y\!\to\! \Jpsi K^+ K^-) \!+\! \Gamma(Y\!\to\! \Jpsi \pi^+ \pi^-) < \Gamma_{Y}$
are jointly satisfied for values $\Lambda_Y > 3550$ MeV.

\begin{figure}[H]
\begin{center}
\epsfig{figure=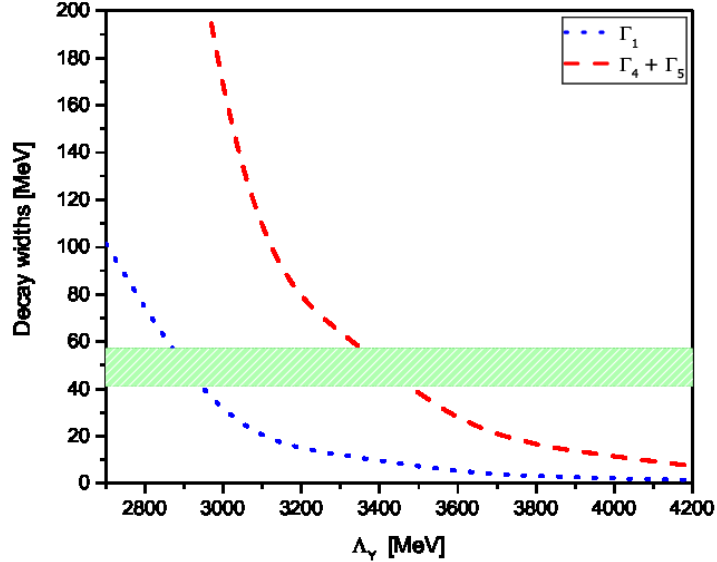,scale=7.0}
\caption{Dependence of the decay widths $\Gamma_1 = \Gamma(Y\to f_0 \Jpsi)$
and $\Gamma(Y\!\to\! \Jpsi K^+ K^-) \!+\! \Gamma(Y\!\to\! \Jpsi \pi^+ \pi^-)$
on size parameter $\Lambda_Y$. The green band depicts the data
$\Gamma_{Y} = 49 \pm 8 \text{MeV}$ ~\cite{PDG2022}.
}
\label{fig:FIG7}
\end{center}
\end{figure}

In final calculations, we use the optimal values of size parameters as follows:
\begin{equation}
\begin{tabular}{ c c c }
\quad $\Lambda_\pi$ \quad & \quad $\Lambda_K$ \quad & \quad $\Lambda_Y$ \quad
\quad
\\\hline
\quad 0.970 GeV \quad & \quad 1.95 GeV \quad & \quad 3.75 GeV \quad
\quad \,.
\end{tabular}
\label{newpara}
\end{equation}

We compare the obtained numerical values of the decay widths and branching
ratio in Table~\ref{tab1} with the most recent experimental data reported in
~\cite{PDG2022} and  ~\cite{Ablikim:2022}.

The most recent experimental data reported in ~\cite{PDG2022} and
~\cite{Ablikim:2022} are reasonably compatible with our estimates of the
Y(4230)-strong decay widths. We observe that the NWA provides rough results
that are overly suppressed,  whereas its modification, PSINWA, provides
more appropriate $Y \!\to\! K^{+} K^{-}\Jpsi$ and $Y \!\to\! \pi^{+} \pi^{-}\Jpsi$
decay widths.

\begin{table}[H]
\centering
\caption{
Our calculations on the strong decay widths of the charmonium-like state
$Y(4230)$ in comparison with the latest data.
}
\label{tab1}
\begin{tabular}{ccccc}
\hline
Decay &
CCQM &
PDG2022 \cite{PDG2022}  &
BESIII \cite{Ablikim:2022}    \\
\hline
$\Gamma(Y \to J/\psi \, f_0$) &
6.9 $\pm$ 0.5 MeV \ &
$\Gamma_Y$ = 49 $\pm$ 8 MeV &
     -      \\
\hline
$\Gamma(f_0 \!\to\! K^{+} K^{-})$&
3.4 $\pm$ 0.3 MeV \ &
$ \Gamma_{f_0}$ = 10 $\div$ 100 MeV &
     -      \\
\hline
$\Gamma(f_0 \!\to\! \pi^{+} \pi^{-})$&
75 $\pm$ 6 MeV \ &
$ \Gamma_{f_0}$ = 10 $\div$ 100 MeV &
     -      \\
\hline
$\Gamma_{NWA}(Y \!\to\! K^{+} K^{-}\Jpsi)$ &
0.29 $\pm$ 0.03 MeV \ &
$ \Gamma_Y$ = 49 $\pm$ 8 MeV &
     -      \\
\hline
$\Gamma_{PSINWA}(Y \!\to\! K^{+} K^{-}\Jpsi)$ &
6.6   $\pm$ 0.5 MeV \ &
$ \Gamma_Y$ = 49 $\pm$ 8 MeV &
   -      \\
\hline
$\Gamma_{integ}(Y \!\to\! K^{+} K^{-}\Jpsi)$ &
7.4 $\pm$ 0.6 MeV \ &
$ \Gamma_Y$ = 49 $\pm$ 8 MeV &
     -      \\
\hline
$\Gamma_{NWA}(Y \!\to\! \pi^{+} \pi^{-}\Jpsi)$ &
6.5 $\pm$ 0.5 MeV \ &
$ \Gamma_Y$ = 49 $\pm$ 8 MeV &
     -      \\
\hline
$\Gamma_{PSINWA}(Y \!\to\! \pi^{+} \pi^{-}\Jpsi)$ &
29.7 $\pm$ 2.5 MeV \ &
$ \Gamma_Y$ = 49 $\pm$ 8 MeV &
     -      \\
\hline
$\Gamma_{integ}(Y \!\to\! \pi^{+} \pi^{-}\Jpsi)$ &
31.0 $\pm$ 2.8 MeV \ &
$ \Gamma_Y$ = 49 $\pm$ 8 MeV &
     -      \\
\hline
$
\left[ \frac{{\cal B}(Y \!\to\! K^{+} K^{-}\Jpsi)}{{\cal B}(Y \!\to\! \pi^{+} \pi^{-}\Jpsi)} \right]_{NWA}
$ &
0.045 $\pm$ 0.012 &
-  &
 0.02 $\div$ 0.26\\
\hline
$
\left[ \frac{{\cal B}(Y \!\to\! K^{+} K^{-}\Jpsi)}{{\cal B}(Y \!\to\! \pi^{+} \pi^{-}\Jpsi)} \right]_{PSINWA}
$ &
0.221  $\pm$ 0.018 &
-  &
 0.02 $\div$ 0.26\\
\hline
$
\left[ \frac{{\cal B}(Y \!\to\! K^{+} K^{-}\Jpsi)}{{\cal B}(Y \!\to\! \pi^{+} \pi^{-}\Jpsi)} \right]_{integ}
$ &
0.236 $\pm$ 0.013  &
-  &
 0.02 $\div$ 0.26\\
\hline
\end{tabular}
\end{table}

According to our estimate shown in Table I, the decay of the resonance
$f_0(980)$ into the $K^{+} K^{-}$ pair is more strongly suppressed compared
to that into the $\pi^{+} \pi^{-}$ pair, while the branching ratio
$|{\cal B}(Y \!\to\! K^{+} K^{-}\Jpsi)/{\cal B}(Y \!\to\! \pi^{+} \pi^{-}\Jpsi)|
_{integ}\simeq 0.236$  is near the experimental upper bound reported
in ~\cite{Ablikim:2022}.

We conclude that the estimated branching ratio and calculated fractal strong decay
widths of the $Y(4230)$ state do not conflict with the latest experimental data,
favoring the molecular representation of the Y-state when the CCQM model is used.



\end{document}